\documentclass[12pt]{article}
\usepackage[margin=1in]{geometry}
\usepackage{amsmath}
\usepackage{amsfonts}
\usepackage{graphicx,psfrag,epsf}
\usepackage{enumerate}
\usepackage{natbib}
\usepackage{url} 
\usepackage{prodint}
\usepackage{amsthm}
\usepackage{multirow}

\newtheorem{thm}{Theorem}
\newtheorem{remark}{Remark}
\newtheorem{corollary}{Corollary}

\newcommand{\blind}{1}
\newcommand{\GG}[1]{}

\begin{document}

\def\spacingset#1{\renewcommand{\baselinestretch}%
{#1}\small\normalsize} \spacingset{1}


\if1\blind
{
  \title{\bf Nonparametric tests for transition probabilities in nonhomogeneous Markov processes}
  \author{Giorgos Bakoyannis\hspace{.2cm}\\
    Department of Biostatistics, Indiana University}
  \maketitle
} \fi

\if0\blind
{
  \bigskip
  \bigskip
  \bigskip
  \begin{center}
    {\LARGE\bf Nonparametric tests for transition probabilities in nonhomogeneous Markov processes}
\end{center}
  \medskip
} \fi

\bigskip
\spacingset{2} 
\begin{abstract}
This paper proposes nonparametric two-sample tests for the direct comparison of the probabilities of a particular transition between states of a continuous time nonhomogeneous Markov process with a finite state space. The proposed tests are a linear nonparametric test, an $L^2$-norm-based test and a Kolmogorov--Smirnov-type test. Significance level assessment is based on rigorous procedures, which are justified through the use of modern empirical process theory. Moreover, the  $L^2$-norm and the Kolmogorov--Smirnov-type tests are shown to be consistent for every fixed alternative hypothesis. The proposed tests are also extended to more complex situations such as cases with incompletely observed absorbing states and non-Markov processes. Simulation studies show that the test statistics perform well even with small sample sizes. Finally, the proposed tests are applied to data on the treatment of early breast cancer from the European Organization for Research and Treatment of Cancer (EORTC) trial 10854, under an illness-death model.
\end{abstract}

\noindent%
\vfill

\section{Introduction}
\label{intro}
Continuous time nonhomogeneous Markov processes with a finite state space are important in many areas of science and particularly in medicine and public health \citep{Tattar14,Bakoyannis19}. Consideration of specific transitions between two states of a multi-state process can provide a deeper and more detailed insight about the treatment effect in clincal trials compared to the analysis of standard survival outcomes, such as event-free survival \citep{Le18}. Important special cases of a Markov process are the univariate survival model, the competing risks model, and the Markov illness-death model \citep{Andersen12}. 

The stochastic behaviour of a Markov process can be described by either the transition intensities, which represent the instantaneous rates of transition between two states, or the transition probabilities. The transition probabilities are also known as survival functions in the framework of the univariate survival model, and as cumulative incidence functions in the competing risks model. It is important to note that, in general, a difference in the transition intensities between two groups does not necessarily imply a difference in the corresponding transition probabilities and vice versa. This phenomenon has been well documented for the special case of the competing risks model \citep{Gray88,Pepe91,Putter07,Bakoyannis12}. Nonparametric tests for comparing transition intesities between groups in general Markov multi-state processes have been well developed \citep{Andersen12}. However, the issue of nonparametric comparison of transition probabilities in general Markov multi-state processes has not received much attention. Nevertheless, transition probabilities, unlike transition intensities, directly quantify clinical prognosis \citep{Bakoyannis19}, which is the target of scientific interest in many applications.

Nonparametric estimation of the transition probabilities of a general Markov process can be performed using the Aalen--Johansen estimator \citep{Aalen78}. The issue of nonparametric comparison of transition probabilities under the univariate survival model has be extensively studied in the literature. For a review of these methods see \citet{Kalbfleisch11} and \citet{Andersen12}. A number of researchers have proposed nonparametric tests for the comparison of transition probabilities for the special case of the competing risks model \citep{Gray88,Pepe93,Lin97}. \citet{Dabrowska00} proposed a graphical procedure based on simultaneous confidence bands to test for differences between transition probabilities in a general Markov process. However, their method imposes proportional hazards assumptions for the transition intensities and, thus, it is not fully nonparametric. Also, this approach does not provide the actual level of statistical significance. \citet{Tattar14} proposed two nonparametric tests for the comparison of the whole transition probability matrices between $k$ groups, by comparing all the possible transition intensities. The first test only compares the transition probability matrices at a specific time point $t_0$, while the second test is a Kolmogorov--Smirnov-type test based on the supremum norm. However, the tests proposed by \citet{Tattar14} do not provide a direct comparison of the transition probability of a particular transition, which is frequently of scientific interest \citep{Le18}. A statistically significant result with these tests only indicates a difference in \textit{any} transition between groups. Recently, \citet{Bluhmki18} proposed a wild bootstrap approach for the Aalen--Johansen estimator, which can be used to construct a simultaneous confidence band for the difference between the transition probabilities of two independent groups. This approach, which is related to a Kolmogorov--Smirnov-type test, can be used as a graphical two-sample comparison procedure at a predetermined $\alpha$ level. However, this approach does not provide the actual level of statistical significance and, also, a Komogorov--Smirnov-type test may not be the most powerful nonparametric test for every alterantive hypotheses. Additionally, there is no rigorous justification about the consistency of this graphical hypothesis testing procedure against any fixed alternative hypothesis \citep{Van00}. Last but not least, the proposed approach is not readily adaptable to more complex situations such as cases with missing data.

This paper addresses the issue of direct nonparametric two-sample comparison of the transition probabilities of a particular transition in a general continuous-time nonhomogeneous Markov process with a finite state space. For this, we propose a linear nonparametric test, an $L^2$-norm-based test and a Kolmogorov--Smirnov-type test. The asymptotic null distributions of the tests are derived. The evaluation of the actual level of statistical significance is based on rigorous procedures justified through the use of modern empirical process theory. Moreover, the $L^2$-norm-based and Kolmogorov--Smirnov-type tests are shown to be consistent against any fixed alternative hypothesis \citep{Van00}. We also propose extensions related to interesting partical problems such as cases with missing absorbing states \citep{Bakoyannis19} and non-Markov processes \citep{Putter18}. The proposed tests exhibit good small sample properties as illustrated in our simulation experiments. Finally, the tests are applied to data on the treatment of early breast cancer from the European Organization for Research and Treatment of Cancer (EORTC) trial 10854.

Compared to the previous work by \citet{Bluhmki18}, which used counting process theory arguments in their derivations, we justify the properties of the proposed tests through the use of modern empirical process theory \citep{Van96,Kosorok08}. As it will be argued later in the text, the practical advantage of our derivations lies on the fact that our proposed tests can be straightforwardly adapted to more complex settings such as cases with incompletely observed absorbing states \citep{Bakoyannis19}. This can be done by replacing the influence function of the standard Aalen--Johansen estimator with the influence function of any other well-behaved and asymptotically linear estimator of the transition probabilities in our proposed testing procedures. Such adaptations are not trivial within the framework of the graphical testing procedure proposed by \citet{Bluhmki18}. An important reason for this is that with more complex estimators, certain predictability conditions assumed by counting process and martingale theory techniques are violated. For such situations, empirical process theory provides a powerful alternative tool. Moreover, we provide two additional tests, a linear test and an $L^2$-norm-based test, which may be more powerful compared to a Kolmogorov--Smirnov-type test in certain settings. Additionally, we argue about the consistency of our $L^2$-norm-based and Kolmogorov--Smirnov-type tests against any fixed alternative hypothesis. Finally, our tests provide the actual level of statistical significance which is useful in pactical applications.

The structure of this paper is as follows. In Section 2 we introduce some notation about Markov processes, provide the proposed nonparametric tests, and consider extensions to more complex situations that are frequently met in practice.  Section 3 presents a simulation study to evaluate the small sample performance  of the proposed tests. Section 4 illustrates the use of the proposed tets using data from the EORTC trial 10854. Finally, Section 5 conlcudes the article with some key remarks. Outlines of the asymptotic theory proofs are provided in the Appendix.

\section{Two-sample nonparametric tests}

\subsection{Nonparametric estimation of transition probabilities}
The stochastic behaviour of a Markov process $\{X(t):t\geq 0\}$ with a finite state space $\mathcal{I}=\{1,\ldots,q\}$ can be described by the $q\times q$ transition probability matrix $\mathbf{P}_0(s,t)=(P_{hj}(s,t))$ whose elements are the transition probabilities
\begin{eqnarray}
P_{hj}(s,t)&=&\Pr(X(t)=j|X(s)=h,\mathcal{F}_{s^-}) \nonumber \\
&=&\Pr(X(t)=j|X(s)=h) \ \ \ \ h,j\in\mathcal{I}, \ \ 0\leq s< t\leq \tau \nonumber 
\end{eqnarray}
where $\mathcal{F}_{s^-}=\sigma\big\langle\{N_{hj}(u):0\leq u< s,h\neq j\}\big\rangle$ is the event history prior to time $s$, with $N_{hj}(t)$ being the number of direct transitions from state $h$ to state $j$, $h\neq j$, in $[0,t]$, $\tau=\sup\{t:\int_0^ta_{hj}(u)du\equiv A_{hj}(t)<\infty,h\neq j\}$, and $a_{hj}(t)=\lim_{h\downarrow 0}P_{hj}(t,t+h)/h$, $h\neq j$, is the transition intensity at time $t$. The conditional independence between the probability of $X(t)$ and the prior history $\mathcal{F}_{s^-}$, conditionally on $X(s)$, is the so-called Markov assumption. Because $\mathbf{P}_0(s,t)$ is a stochastic matrix we have that $a_{hh}(t)=-\sum_{j\neq h} a_{hj}(t)$.

The observed data from a sample of i.i.d. observations of a Markov process are the counting processes $\{N_{ihj}(t):h\neq j,t\in[0,\tau]\}$, $i=1,\ldots,n$, which represent the number of direct transitions of the $i$th observation from the state $h$ to the state $j$ by time $t$, and the at-risk processes $\{Y_{ih}(t):h\in
\mathcal{I},t\in[0,\tau]\}$ which are the indicator functions of whether the $i$th observation is at the state $h\in\mathcal{I}$ just before time $t\in[0,\tau]$. Based on such a sample, the transition probability matrix of a nonhomogeneous Markov process can be estimated using the Aalen--Johansen estimator \citep{Aalen78}:
\[
\hat{\mathbf{P}}_n(s,t)=\Prodi_{(s,t]}\left[\mathbf{I}+d\hat{\mathbf{A}}_n(u)\right], \ \ \ \ s,t\in[0,\tau],
\]
where $\prodi$ is the product integral and $\hat{\mathbf{A}}_n(t)$ a $q\times q$ matrix whose elements are the Nelson--Aalen estimates of the cumulative transition intensities
\[
\hat{A}_{n,hj}(t)=\int_0^t\frac{\sum_{i=1}^n{dN_{ihj}(u)}}{\sum_{i=1}^nY_{ih}(u)}, \ \ \ \ h\neq j.
\]

\subsection{Linear nonparametric tests}
\label{linear}

First consider the two-sample problem of comparing the transition probabilities $P_{0,hj}^{(1)}(s,\cdot)$ and $P_{0,hj}^{(2)}(s,\cdot)$, $s\in[0,\tau)$, of two populations of interest, for a particular transition $h\rightarrow j$, with $h,j\in\mathcal{I}$. For simplicity of presentation we will set the starting point $s=0$ for the remainder of the paper. Based on two independent random samples of $n_1$ and $n_2$ observations from the two populations, define the pointwise weighted difference
\[
D_{hj}(t)=\hat{W}_{hj}(t)\left[\hat{P}_{n_1,hj}^{(1)}(0,t)-\hat{P}_{n_2,hj}^{(2)}(0,t)\right], \ \ \ \ t\in[0,\tau]
\]
where $\hat{W}_{hj}(t)$ is a weight function and $\hat{P}_{n_1,hj}^{(1)}(0,t)$ and $\hat{P}_{n_2,hj}^{(2)}(0,t)$ are the nonparametric Aalen--Johansen estimates of the transition probabilities of the two populations under comparison. Example of weight function choices are $W_{hj}(t)=1$ and
\begin{equation*}
\hat{W}_{hj}(t)=\frac{\bar{Y}_h^{(1)}(t)\bar{Y}_h^{(2)}(t)}{\bar{Y}_h^{(1)}(t)+\bar{Y}_h^{(2)}(t)}
\end{equation*}
where $\bar{Y}_h^{(p)}(t)=n_p^{-1}\sum_{i=1}^{n_p}Y_{ih}^{(p)}(t)$, $p=1,2$. The latter choice assigns more weight to times with more observations at risk. A natural linear test for the null hypothesis $H_0:P_{0,hj}^{(1)}=P_{0,hj}^{(2)}$ is the area under the weighted difference curve 
\[
Z_{hj}=\int_{(0,\tau]}D_{hj}(t)dm(t),
\]
where $m$ is the Lebesgue measure on the Borel $\sigma$-algebra $\mathcal{B}([0,\tau])$. To establish the asymptotic distribution of the test statistic $Z_{hj}$ we assume the following conditions.
\begin{itemize}
\item[\textbf{C1.}] The potential right censoring and left truncation are independent of the counting processes $\{N_{hj}(t):h\neq j, t\in[0,\tau]\}$ and noninformative about $\mathbf{P}_0(s,t)$.
\item[\textbf{C2.}] $n_1/(n_1+n_2)\rightarrow \lambda\in(0,1)$ as $\min(n_1,n_2)\rightarrow\infty$.
\item[\textbf{C3.}] The counting processes $\{N_{hj}(t):h\neq j, t\in[0,\tau]\}$ are bounded in the sense that $\Pr(N_{hj}(\tau)\leq C)=1$ for some constant $C\in (0,\infty)$.
\item[\textbf{C4.}] $\inf_{t\in[0,\tau]}E[Y_h(t)]>0$ for all the transient states $h\in\mathcal{I}$.
\item[\textbf{C5.}] The cumulative transition intensities $\{A_{0,hj}(t):h\neq j, t\in[0,\tau]\}$ are continuous functions of bounded variation on $[0,\tau]$.
\item[\textbf{C6.}] The weight $\hat{W}_{hj}(t)$ converges uniformly to a nonnegative uniformly bounded function $W_{hj}(t)$ on $[0,\tau]$.
\end{itemize}
\begin{remark}
\normalfont In some applications condition C4 may not be satisfied for some timepoints for one or more states $h\in\mathcal{I}$. In such cases one can restrict the comparison interval to $[t_1,t_2]$ with $0<t_1<t_2<\tau$, such that $\inf_{t\in[t_1,t_2]}E[Y_h(t)]>0$ for those $h\in\mathcal{I}$. In such cases the test statistic becomes
\[
Z_{hj}=\int_{(t_1,t_2]}D_{hj}(t)dm(t).
\]
\end{remark}

Before stating the theorem about the asymptotic distribution of test statistic we define the functions
\[
M_{ilm}^{(p)}(t)=N_{ilm}^{(p)}(t)-\int_{(0,t]}Y_{il}^{(p)}(u)dA_{0,lm}^{(p)}(u),
\]
where $N_{ilm}^{(p)}(t)$ and $Y_{il}^{(p)}(t)$ are the counting and at-risk processes of the $i$th observation in the $p$th sample at time $t$. Also, define $\mathcal{T}$ to be the subset of $\mathcal{I}$ which contains the potential absorbing states. The set $\mathcal{T}$ will be null for non-absorbing Markov processes.

Theorem 1 provides the asymptotic distribution of $Z_{hj}$ under the null hypothesis.

\begin{thm}
Suppose that conditions C1-C6 hold. Then under the null hypothesis 
\[
\sqrt{\frac{n_1n_2}{n_1+n_2}}Z_{hj}\overset{d}\rightarrow G_{hj},
\]
where $G_{hj}\sim N(0,\omega_{hj}^2)$ and
\[
\omega_{hj}^2=(1-\lambda)E\left[\int_{(0,\tau]}W_{hj}(t)\gamma_{1hj}^{(1)}(0,t)dm(t)\right]^2+\lambda E\left[\int_{(0,\tau]}W_{hj}(t)\gamma_{1hj}^{(2)}(0,t)dm(t)\right]^2.
\]
with 
\[
\gamma_{ihj}^{(p)}(s,t)=\sum_{l\notin\mathcal{T}}\sum_{m\in\mathcal{I}}\int_s^t\frac{P_{0,hl}^{(p)}(s,u-)P_{0,mj}^{(p)}(u,t)}{EY_{1l}^{(p)}(u)}dM_{ilm}^{(p)}(u), \ \ \ \ 0\leq s<t \leq \tau, \ \ p=1,2, 
\]
for $i=1,\ldots,n_p$. 
\end{thm}

\begin{remark}
\normalfont The functions $\gamma_{ihj}^{(p)}(s,t)$, p=1,2, in Theorem 1 are the influence functions of the Aalen--Johansen estimator.
\end{remark}
A consistent (in probability) estimator of the variance $\omega_{hj}^2$ is 
\begin{eqnarray*}
\hat{\omega}_{hj}^2&=&\frac{n_2}{(n_1+n_2)n_1}\sum_{i=1}^{n_1}\left[\int_{(0,\tau]}\hat{W}_{hj}(t)\hat{\gamma}_{ihj}^{(1)}(0,t)dm(t)\right]^2 \\
&&+\frac{n_1}{(n_1+n_2)n_2}\sum_{i=1}^{n_2}\left[\int_{(0,\tau]}\hat{W}_{hj}(t)\hat{\gamma}_{ihj}^{(2)}(0,t)dm(t)\right]^2,
\end{eqnarray*}
where $\gamma_{ihj}^{(p)}(0,t)$, $p=1,2$, are estimated by replacing the expectations with sample averages and the unknown parameters with their uniform consistent estimates. Now, Theorem 1 and $\hat{\omega}_{hj}$ can be used to constuct a $Z$-test for the null hypothesis as:
\[
\frac{Z_{hj}}{\hat{\omega}_{hj}\Big/\sqrt{\frac{n_1n_2}{n_1+n_2}}}.
\]
The actual significance level can then be avaluated under the standard normal distribution as usual.

\subsection{$L^2$-norm-based and Kolmogorov--Smirnov-type tests}
\label{omnibus}

A linear test is not the optimal choice when the two transition probability curves under comparison cross at one or more timepoints. In this section, we propose alternative tests for such situations. The first test is a test based on an $L^2$ norm
\[
Q_{1hj}=\left\{\int_{(0,\tau]}\left[D_{hj}(t)\right]^2dm(t)
\right\}^{1/2}
\]
while the second test is a Kolmogorov--Smirnov-type test
\[
Q_{2hj}=\sup_{[0,\tau]}|D_{hj}(t)|.
\]
The Kolmogorov--Smirnov-type test is related to the graphical hypothesis testing procedure proposed by \citet{Bluhmki18}. The asymptotic null distributions of these tests are complicated. However, significance level can be easily calculated numerically by proper simulation realizations from the null distribution of these test statistics. Theorem 2 provides the basis for an approach to properly simulate realizations from the null distributions of $Q_{1hj}$ and $Q_{2hj}$. Before stating Theorem 2 define the estimated functions
\begin{eqnarray*}
\hat{B}_{hj}(t)&=&\sqrt{1-\lambda}\frac{1}{\sqrt{n_1}}\sum_{i=1}^{n_1}\hat{W}_{hj}(t)\hat{\gamma}_{ihj}^{(1)}(0,t)\xi_{i}^{(1)} \\
&&-\sqrt{\lambda}\frac{1}{\sqrt{n_2}}\sum_{i=1}^{n_2}\hat{W}_{hj}(t)\hat{\gamma}_{ihj}^{(2)}(0,t)\xi_{i}^{(2)}, \ \ \ \ h,j\in\mathcal{I},\ \ t\in[0,\tau]
\end{eqnarray*}
where $\{\xi_{i}^{(p)}\}_{i=1}^{n_p}$, $p=1,2$, are independent draws from $N(0,1)$. 
\begin{thm}
Suppose that conditions C1-C6 hold. Then under the null hypothesis 
\[
\sqrt{\frac{n_1n_2}{n_1+n_2}}D_{hj}\leadsto\sqrt{1-\lambda}\mathbb{G}_{1hj}-\sqrt{\lambda}\mathbb{G}_{2hj},
\]
and, conditionally on the observed data,
\[
\hat{B}_{hj}\leadsto\sqrt{1-\lambda}\mathbb{G}_{1hj}-\sqrt{\lambda}\mathbb{G}_{2hj},
\]
where $\mathbb{G}_{1hj}$ and $\mathbb{G}_{2hj}$ are two independent tight zero-mean Gaussian processes with covariance functions
\[
\sigma_{hjp}(v,t)=E[W_{hj}(v)\gamma_{1hj}^{(p)}(0,v)][W_{hj}(t)\gamma_{1hj}^{(p)}(0,t)], \ \ \ \ p=1,2.
\]
\end{thm}

\begin{corollary}
\normalfont By Theorem 2 and the continuous mapping theorem it follows that under the null hypothesis
\[
\sqrt{\frac{n_1n_2}{n_1+n_2}}Q_{1hj}\overset{d}\rightarrow \left\{\int_{(0,\tau]}\left[\sqrt{1-\lambda}\mathbb{G}_{1hj}(t)-\sqrt{\lambda}\mathbb{G}_{2hj}(t)\right]^2dm(t)\right\}^{1/2},
\]
and
\[
\sqrt{\frac{n_1n_2}{n_1+n_2}}Q_{2hj}\overset{d}\rightarrow  \sup_{t\in[0,\tau]}\left|\sqrt{1-\lambda}\mathbb{G}_{1hj}(t)-\sqrt{\lambda}\mathbb{G}_{2hj}(t)\right|.
\]
\end{corollary}

The asymptotic null distributions of the omnibus tests are quite complicated and, thus, they are of limited use in terms of evaluating the significance level. However, Theorem 2 provides justification about a way to numerically calculate $p$-values through a simple simulation technique. This can be performed as follows. In light of Theorem 2, one can simulate from the asymptotic null asymptotic distributions of the tests $Q_{1hj}$ and $Q_{2hj}$ by simulating multiple versions of $\{\xi_{ir}^{(1)}\}_{i=1}^{n_1}$ and $\{\xi_{ir}^{(2)}\}_{i=1}^{n_2}$  independently from $N(0,1)$ for $r=1,\ldots,R$, and then calculating a sample for the above null distributions as
\[
\left\{\int_{(0,\tau]}\left[\hat{B}_{hj,r}(t)\right]^2dm(t)\right\}^{1/2}, \ \ \ \ r=1,\ldots,R
\]
and $\sup_{t\in[0,\tau]}\left|\hat{B}_{hj,r}(t)\right|$, $ r=1,\ldots,R$, respectively, where
\begin{eqnarray*}
\hat{B}_{hj,r}(t)&=&\sqrt{1-\lambda}\frac{1}{\sqrt{n_1}}\sum_{i=1}^{n_1}\hat{W}_{hj}(t)\hat{\gamma}_{ihj}^{(1)}(0,t)\xi_{ir}^{(1)} \\
&&-\sqrt{\lambda}\frac{1}{\sqrt{n_2}}\sum_{i=1}^{n_2}\hat{W}_{hj}(t)\hat{\gamma}_{ihj}^{(2)}(0,t)\xi_{ir}^{(2)}, \ \ \ \ r=1,\ldots,R.
\end{eqnarray*}
Now, the significance level for each test can be calculated as the proportion of realizations from the corresponding null distribution that is greater than or equal to the calculated tests statistic value from the observed data.

The tests $Q_{1hj}$ and $Q_{2hj}$ are consistent for every fixed alternative hypothesis with $P_{0,hj}^{(1)}\neq P_{0,hj}^{(2)}$. This follows from Theorem 2, the uniform consistency of the Aalen--Johansen estimator of the transition probabilities \citep{Aalen78}, condition C6, the continuity of these tests in $D_{hj}(t)$, and Lemma 14.15 in \citet{Van00}.

\subsection{Extensions to more complex settings}

Many complications that frequently occur in practice make the application of the proposed tests improper. An important example is the problem of incompletely observed absorbing states, where missingness occurs either due to the usual nonresponse or the study design \citep{Bakoyannis19}. A special case of this is the issue of missing causes of death in biomedical applications. In such cases, a complete case analysis, which discards cases with a missing cause of death, is well known to lead to biased estimates \citep{Gao05,Lu08,Bakoyannis19}. In general, more complicated cases require extensions of the standard Aalen--Johansen estimator, denoted by $\tilde{P}_{n,hj}(s,t)$, to consistently estimate the transition probabilities of interest over a compact interval $H\subset [0,\tau]$. In such cases, one can replace the standard Aalen--Johansen estimator with another approriate estimator $\tilde{P}_{n,hj}(s,t)$ in the testing procedures. Then, the linear test becomes 
\[
\tilde{Z}_{hj}=\int_{H}\tilde{D}_{hj}(t)dm(t),
\]
where
\[
\tilde{D}_{hj}(t)=\hat{W}_{hj}(t)\left[\tilde{P}_{n_1,hj}^{(1)}(s,t)-\tilde{P}_{n_2,hj}^{(2)}(s,t)\right], \ \ \ \ t\in H,
\]
while the $L^2$-norm based and Kolmogorov--Smirnov-type tests become
\[
\tilde{Q}_{1hj}=\left\{\int_{H}\left[\tilde{D}_{hj}(t)\right]^2dm(t)
\right\}^{1/2}
\]
and
\[
\tilde{Q}_{2hj}=\sup_{t\in H}|\tilde{D}_{hj}(t)|.
\]

The following conditions ensure the validity of the proposed testing procedures in more complex settings.

\begin{itemize}
\item[\textbf{D1.}] The estimator $\tilde{P}_{n,hj}(s,\cdot)$ is consistent in the sense
\[
\sup_{t\in H}|\tilde{P}_{n,hj}(s,t)-P_{hj}(s,t)|\overset{p}\rightarrow 0,
\] 
for some $s\geq 0$, where $H$ is a compact subset of $[0,\tau]$.
\item[\textbf{D2.}] The estimator $\tilde{P}_{n,hj}(s,\cdot)$ is an asymptotically linear estimator with
\[
\sqrt{n}[\tilde{P}_{n,hj}(s,t)-P_{hj}(s,t)]=\frac{1}{\sqrt{n}}\sum_{i=1}^n\phi_{ihj}(s,t)+o_p(1),
\]
where the influence functions $\phi_{ihj}(s,t)$ belong to a Donsker class indexed by $H$.
\item[\textbf{D3.}] The empirical versions of the influence functions $\hat{\phi}_{ihj}(s,t)$ satisfy
\[
\sup_{t\in H}\left|n^{-1/2}\sum_{i=1}^n[\hat{\phi}_{ihj}(s,t)-\phi_{ihj}(s,t)]\xi_i\right|\overset{p}\rightarrow 0,
\]
where $\xi_i$ are independent random draws from $N(0,1)$.
\end{itemize}

\begin{remark}
\normalfont Condition D2 is sufficient for establishing the weak convergence of the estimator $\tilde{P}_{n,hj}(s,\cdot)$ to a tight mean-zero Gaussian process. Condition D3 along with the conditional multiplier central limit theorem \citep{Van96,Kosorok08} and condition D2, provide a simulation approach for the construction of simultaneous confidence bands \citep{Kosorok08}. Therefore, conditions D1-D3 are expected to have been established in works extending the standard Aalen--Johansen estimator to more complex settings. This is the case, for example, for the nonparametric estimator of the transition probability matrix with incompletely observed absorbing states \citep{Bakoyannis19}.
\end{remark}

Hypothesis testing in more complex settings can be simply performed by replacing the influence functions $\gamma_{ihj}^{(p)}(s,t)$, $p=1,2$, of the standard Aalen--Johansen estimator with the influence functions $\phi_{ihj}^{(p)}(s,t)$ of the estimator $\tilde{P}_{n,hj}(s,t)$. The theorems stated below justify the direct use of the proposed tests in more complex situations. Before stating those theorems define the functions
\begin{eqnarray*}
\tilde{B}_{hj}(t)&=&\sqrt{1-\lambda}\frac{1}{\sqrt{n_1}}\sum_{i=1}^{n_1}\hat{W}_{hj}(t)\hat{\phi}_{ihj}^{(1)}(s,t)\xi_{i}^{(1)} \\
&&-\sqrt{\lambda}\frac{1}{\sqrt{n_2}}\sum_{i=1}^{n_2}\hat{W}_{hj}(t)\hat{\phi}_{ihj}^{(2)}(s,t)\xi_{i}^{(2)}, \ \ \ \ h,j\in\mathcal{I},\ \ t\in H
\end{eqnarray*}
where $\{\xi_{i}^{(p)}\}_{i=1}^{n_p}$, $p=1,2$, are independent draws from $N(0,1)$. 

\begin{thm}
Suppose that conditions C2, C6, D1 and D2 hold. Then under the null hypothesis 
\[
\sqrt{\frac{n_1n_2}{n_1+n_2}}\tilde{Z}_{hj}\overset{d}\rightarrow \tilde{G}_{hj},
\]
where $\tilde{G}_{hj}\sim N(0,\theta_{hj}^2)$ and
\[
\theta_{hj}^2=(1-\lambda)E\left[\int_{H}W_{hj}(t)\phi_{1hj}^{(1)}(s,t)dm(t)\right]^2+\lambda E\left[\int_{H}W_{hj}(t)\phi_{1hj}^{(2)}(s,t)dm(t)\right]^2.
\]
\end{thm}
The proof of Theorem 3 involves the same arguments to those used in the proof of Theorem 1 given in the Appendix.
\begin{thm}
Assume that conditions C2, C6, and D1--D3 are satisfied. Then, under the null hypothesis 
\[
\sqrt{\frac{n_1n_2}{n_1+n_2}}\tilde{D}_{hj}\leadsto\sqrt{1-\lambda}\tilde{\mathbb{G}}_{1hj}-\sqrt{\lambda}\tilde{\mathbb{G}}_{2hj},
\]
and, conditionally on the observed data,
\[
\tilde{B}_{hj}\leadsto\sqrt{1-\lambda}\tilde{\mathbb{G}}_{1hj}-\sqrt{\lambda}\tilde{\mathbb{G}}_{2hj},
\]
where $\tilde{\mathbb{G}}_{1hj}$ and $\tilde{\mathbb{G}}_{2hj}$ are two independent tight zero-mean Gaussian processes with covariance functions
\[
\tilde{\sigma}_{hjp}(v,t)=E[W_{hs}(v)\phi_{1hj}^{(p)}(s,v)][W_{hj}(t)\phi_{1hj}^{(p)}(s,t)], \ \ \ \ p=1,2.
\]
\end{thm}
The proof of Theorem 4 follows from similar arguments to those used in the proof of Theorem 2 given in the Appendix.

\subsubsection{Missing absorbing states}
In many settings one can observe that a process has arrived at some absorbing state, but the actual absorbing state is unobserved for some study participants, such as in cases with missing causes of death. For such situations, \citet{Bakoyannis19} proposed a nonparametric maximum pseudolikelihood estimator (NPMPLE) under a missing at random assumption. To review this estimator, let $\Delta_{ij}$ be an indicator variable with $\Delta_{ij}=1$ if the $i$th observation arrived at the absorbing state $j\in\mathcal{T}$, and $\Delta_{ij}=0$ otherwise. Also, let $R_i$ be another indicator variable with $R_i=1$ indicating that the absorbing state of the $i$th observation has been successfully ascertained. Finally, let $\pi_j(\mathbf{O}_i,\boldsymbol{\beta}_0)$ be the probability that $\Delta_{ij}=1$ given the fully observed data $\mathbf{O}_i$, under a parametric model indexed by an unknown Euclidean parameter $\boldsymbol{\beta}_0$. In this setting, the cumulative transition intensities can be estimated using the NPMPLE:
\[
\tilde{A}_{n,hj}(t)=\int_0^t\frac{\sum_{i=1}^n{d\tilde{N}_{ihj}(u;\hat{\boldsymbol{\beta}}_n)}}{\sum_{i=1}^nY_{ih}(u)}, \ \ \ \ h\neq j, \ \ j\in\mathcal{T},
\]
where
\[
\tilde{N}_{ihj}(t;\hat{\boldsymbol{\beta}}_n)=[R_i\Delta_{ij}+(1-R_{ij})\pi_j(\mathbf{O}_i,\hat{\boldsymbol{\beta}}_n)]\sum_{l\in\mathcal{T}}N_{ihl}(t),
\]
with $\hat{\boldsymbol{\beta}}_n$ being a consistent estimator of $\boldsymbol{\beta}_0$. The transition probability matrix can then be estimated as
\[
\tilde{\mathbf{P}}_n(s,t)=\Prodi_{(s,t]}\left[\mathbf{I}+d\tilde{\mathbf{A}}_n(u)\right], \ \ \ \ s,t\in[0,\tau],
\]
where the components of the matrix $\tilde{\mathbf{A}}_n(u)$ are $\tilde{A}_{n,hj}(u)$. By Theorems 1 and 2 in \citet{Bakoyannis19} and calculations provided in the proof of Theorem 2 in the same source, the NPMPLE estimator satisfies the conditions D1-D3 above. Therefore, if the conditions in \citet{Bakoyannis19} and the conditions C2 and C6 above are satisfied, two-sample comparison can be performed by utilizing the NPMPLE of the transition probabilities along with the corresponding influence functions in the proposed tests. This is justified by Theorems 3 and 4 above.

\subsubsection{Non-Markov processes}

Trivially, the Aalen--Johansen estimator $\hat{P}_{n,hj}(0,\cdot)$ is uniformly consistent for the transition probability $P_{hj}(0,\cdot)$ even under a non-Markov process \citep{Datta01,Titman15}. When the interest lies on the marginal $\Pr(X(t)=j|X(s)=h)$, i.e. unconditionally on the prior history $\mathcal{F}_{s^-}$, for some $s>0$, under a non-Markov process, then the landmark Aalen--Johansen estimator is consistent for $\Pr(X(t)=j|X(s)=h)$ \citep{Putter18} under the conditions of \citet{Datta01} and, also, the assumption that $\Pr(X(s)=h)>0$. The landmark Aalen--Johansen estimator is essentially equivalent to the standard Aalen--Johansen estimator, except for the fact that only observations with $X(s)=h$ are considered. This is achieved by considering the modified counting and at-risk processes $\tilde{N}_{ihj}(t)=N_{ihj}(t)I(X(s)=h)$ and $\tilde{Y}_{ih}(t)=Y_{ih}(t)I(X(s)=h)$, for $t\geq s$. Therefore, the  influence functions of the landmark Aalen--Johansen estimator are the same to that of the standard Aalen--Johansen estimator, with the only exception that the former involves the modified $\tilde{N}_{ihj}(t)$ and $\tilde{Y}_{ih}(t)$ instead of the standard counting and at-risk processes $N_{ihj}(t)$ and $Y_{ih}(t)$. Consequently, it is clear that conditions D1--D3 are satified if $\Pr(X(s)=h)>0$ and, also, if the conditions in \citet{Datta01} hold. Thus, in light of Theorems 3 and 4, the proposed nonparametric tests can be used with non-Markov processes by utilizing the landmark Aalen--Johansen estimator.

\section{Simulation studies}
\label{sims}
\begin{figure}
\centering
\caption{Illness-death model without recovery assumed in the simulation study.}\label{model}
\includegraphics[scale=1]{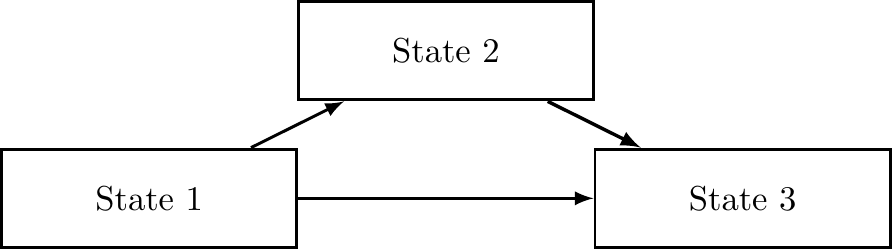}
\end{figure}

To evaluate the finite sample performance of the proposed test statistics, we conducted a simulation study. We considered a nonhomogeneous Markov process with 2 transient states $\{1,2\}$ and 1 absorbing state $\{3\}$, under the illness-death model without recovery \citep{Andersen12}. This model is illustrated in Figure~\ref{model}. In this simulation study, we focused on the null hypothesis $H_0:P_{12}^{(1)}=P_{12}^{(2)}$. Initially, we independently generated the times from state 1 to states 2 and 3 by assuming the cumulative transition intensities $A_{12}(t)=\alpha_1t$ and $A_{13}(t)=t/2$. For observations that first arrived at the transient state 2, we generated the time from state 2 to the absorbing state 3, assuming a cumulative transition intensity $A_{23}(t)=\alpha_2t$. The parameter values considered were $\alpha_{1}\in\{0.4,0.6,1.2,1.4\}$ and $\alpha_{2}\in\{0.25,0.5,0.75\}$. Then, the right censoring times were independently simulated from Exp(0.25). Under this set-up the transition probability of interest was
\[
P_{12}(s,t)=\frac{\alpha_1\left[e^{\alpha_2(s-t)}-e^{(\alpha_1+0.5)(s-t)}\right]}{\alpha_1-\alpha_2+0.5}.
\]
Different simulation scenarios were considered according to the sample sizes $n_p$, $p=1,2,$ and the parameter values of the two groups. 1,000 datasets were simulated for each scenario, and the $L^2$ distance test and Kolmogorov--Smirnov-type test were calculated using 1,000 independent simulations of $\{\xi_{i}^{(1)}\}_{i=1}^{n_1}$ and $\{\xi_{i}^{(2)}\}_{i=1}^{n_2}$ from $N(0,1)$. Finally, the weight function $W_{hj}(t)=1$ was considered in all cases.

Simulation results regarding the empirical type I error rates are presented in Tables~\ref{H0a} and~\ref{H0b}, respectively. Under these scenaria, the empirical type I errors rates for all tests were close to the nominal $\alpha$ levels, even in situations with small sample sizes. Thus, these results provide numerical evidence for the validity of the proposed hypothesis testing procedures under $H_0$. Simulation results regarding the empirical power levels under alternative hypotheses with non-crossing transition probabilities are presented in Table~\ref{H1a}. Under these scenaria, the empirical power levels increased with sample size and, also, with a more pronounced difference between the two groups, as expected. The power levels for the three tests were in general similar. However, under a less pronounced difference between the two groups, the linear test exhibited a somewhat larger empirical power with larger sample sizes. These results provide numerical evidence for the consistency of the proposed tests with non-crossing transition probabilities. Simulation results regarding the empirical power levels under alternative hypotheses with crossing transition probabilities are presented in Table~\ref{H1b}. These scenaria illustrate numerically the inconsistency of the linear tests with crossing transition probabilities, as the empirical power levels did not systematically increase with sample size. On the contrary, the empirical power of the $L^2$-norm-based and Kolmogorov--Smirnov-type tests increased with sample size and with a more pronounced difference between the two groups. These results indicate numerically the consistency of the omnibus tests against alternatives with crossing transition probabilities.

\begin{table}
\caption{Simulation results about empirical type I error rates for the linear test (Linear), the $L^2$-norm-based test ($L^2$), and the Kolmogorov--Smirnov-type test (KS) under simulation scenaria 1 and 2.}\label{H0a}
\begin{tabular}{c c c c c c c c c}
\hline
&&&\multicolumn{3}{c}{$\alpha=0.01$} & \multicolumn{3}{c}{$\alpha=0.05$} \\
$H_0$ scenario & $n_1$ & $n_2$ & Linear & $L^2$ & KS & Linear & $L^2$ & KS \\
\hline
\multirow{5}{*}{\parbox[c]{12em}{\includegraphics[scale=0.27]{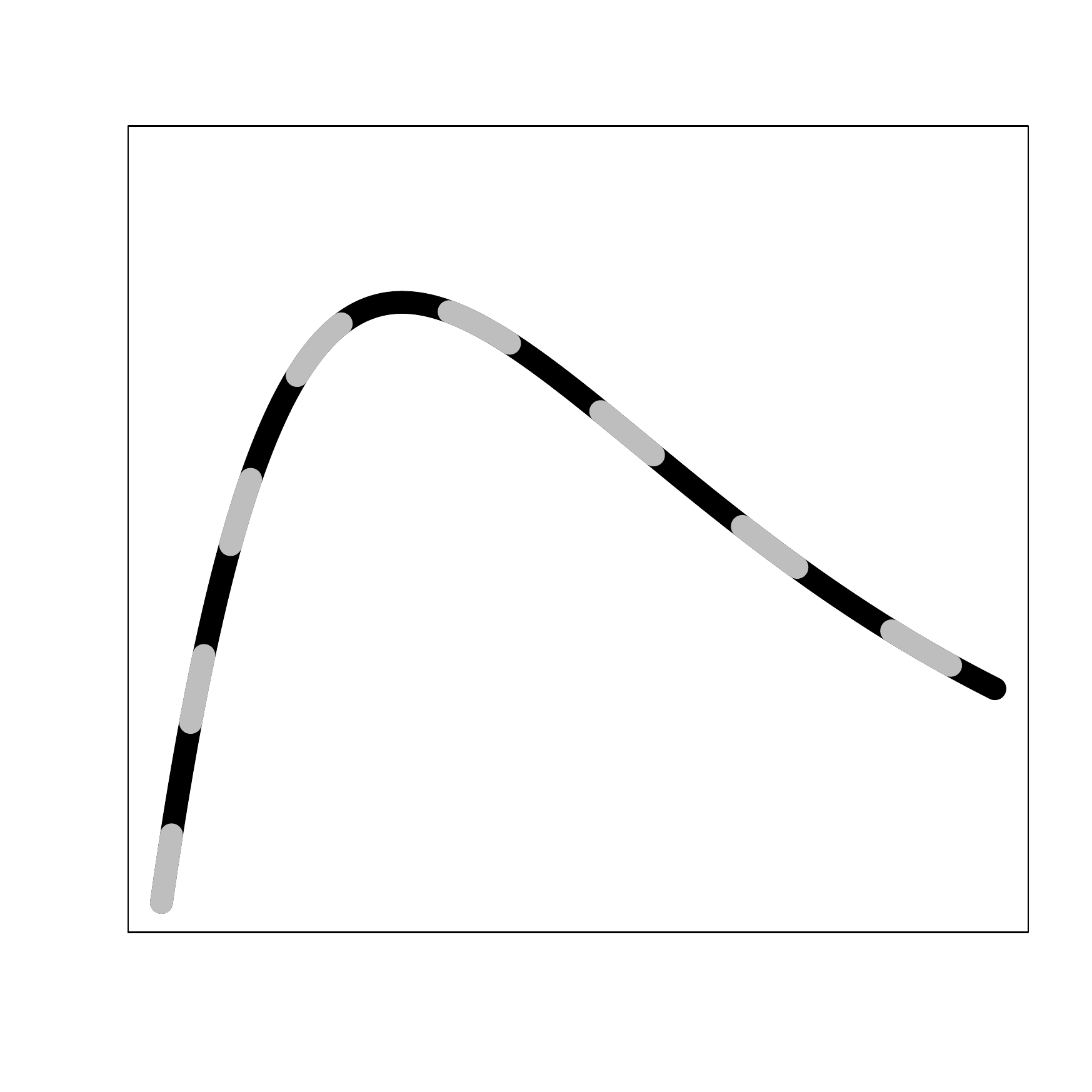}}} & 50 & 50 & 0.009 & 0.007 & 0.015 & 0.051 & 0.054 & 0.063 \\
 & 100 & 50 & 0.010 & 0.009 & 0.018 & 0.053 & 0.051 & 0.047 \\
 & 100 & 100 & 0.012 & 0.011 & 0.012 & 0.060 & 0.060 & 0.054 \\
 & 200 & 100 & 0.013 & 0.014 & 0.008 & 0.049 & 0.047 & 0.045 \\
 & 200 & 200 & 0.010 & 0.012 & 0.010 & 0.051 & 0.047 & 0.052 \\
\hline
\multirow{5}{*}{\parbox[c]{12em}{\includegraphics[scale=0.27]{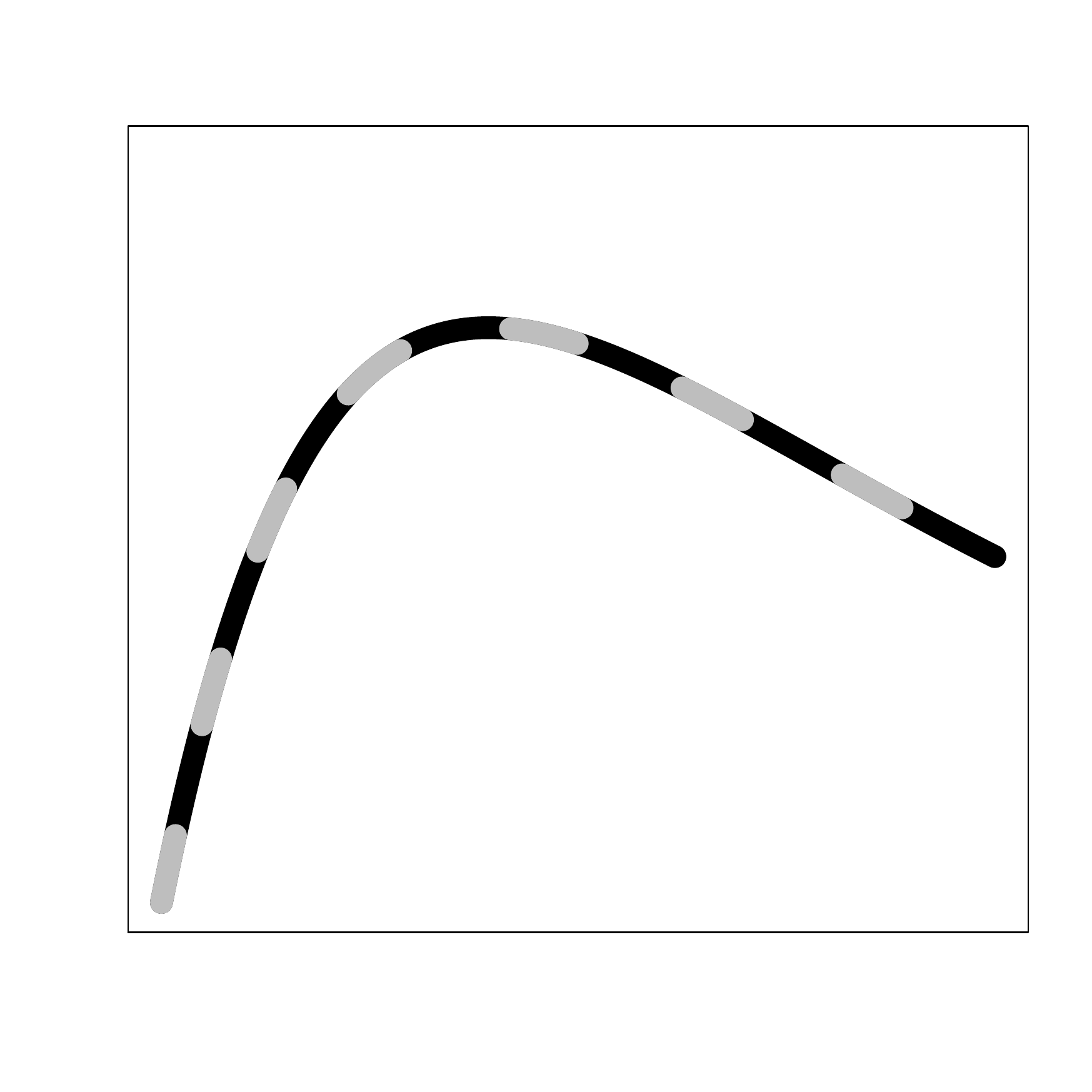}}} & 50 & 50 & 0.014 & 0.013 & 0.015 & 0.067 & 0.061 & 0.066 \\
& 100 & 50 & 0.014 & 0.014 & 0.017 & 0.057 & 0.053 & 0.065 \\
& 100 & 100 & 0.016 & 0.015 & 0.013 & 0.058 & 0.057 & 0.056 \\
& 200 & 100 & 0.011 & 0.014 & 0.009 & 0.058 & 0.060 & 0.057 \\
& 200 & 200 & 0.008 & 0.014 & 0.016 & 0.060 & 0.055 & 0.062 \\
\hline
\end{tabular}
\end{table}

\begin{table}
\caption{Simulation results about empirical type I error rates for the linear test (Linear), the $L^2$-norm-based test ($L^2$), and the Kolmogorov--Smirnov-type test (KS) under simulation scenaria 3 and 4.}\label{H0b}
\begin{tabular}{c c c c c c c c c}
\hline
&&&\multicolumn{3}{c}{$\alpha=0.01$} & \multicolumn{3}{c}{$\alpha=0.05$} \\
$H_0$ scenario & $n_1$ & $n_2$ & Linear & $L^2$ & KS & Linear & $L^2$ & KS \\
\hline
\multirow{5}{*}{\parbox[c]{12em}{\includegraphics[scale=0.27]{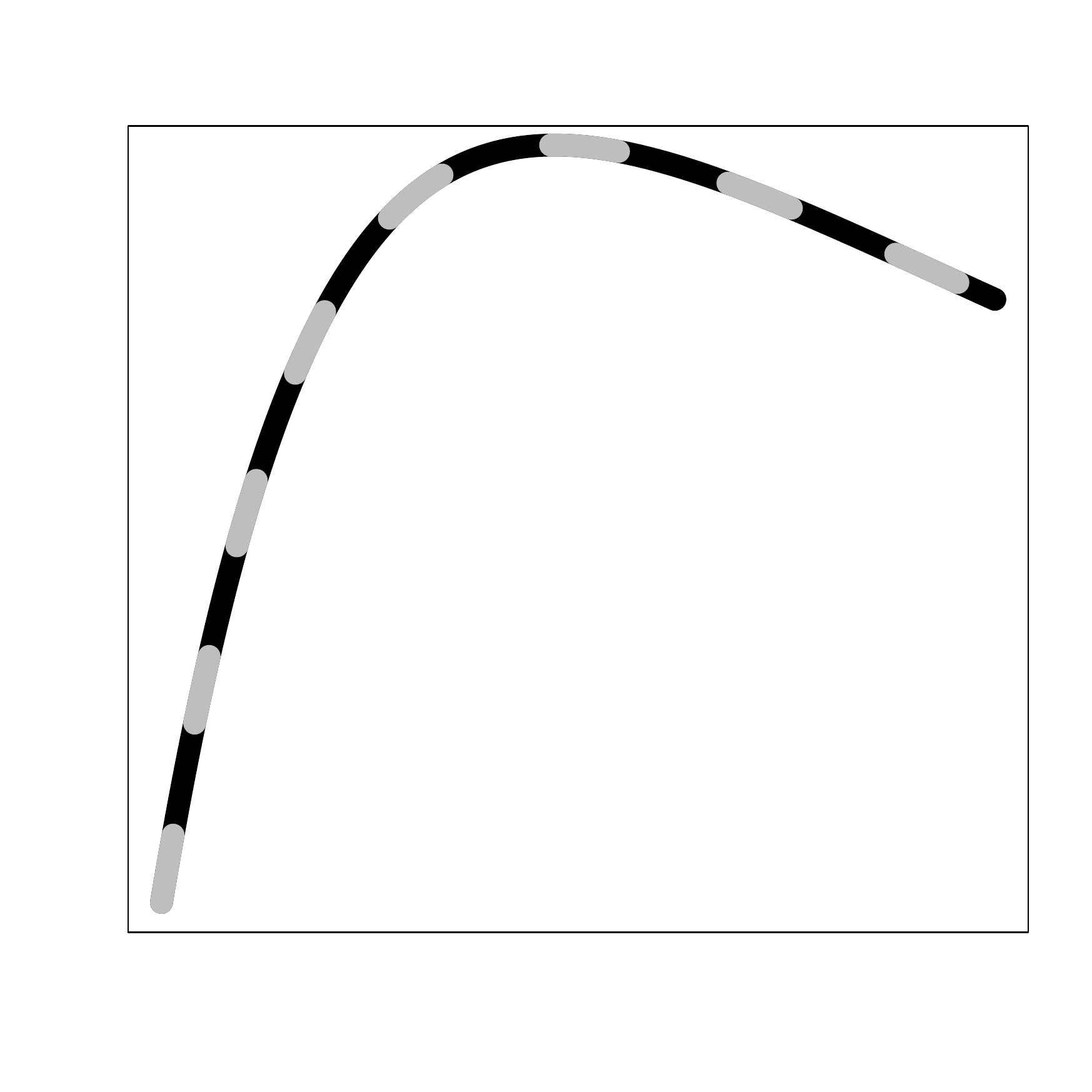}}} & 50 & 50 & 0.016 & 0.014 & 0.011 & 0.075 & 0.069 & 0.066 \\
 & 100 & 50 & 0.014 & 0.018 & 0.013 & 0.052 & 0.048 & 0.060 \\
 & 100 & 100 & 0.011 & 0.012 & 0.012 & 0.048 & 0.047 & 0.046 \\
 & 200 & 100 & 0.009 & 0.009 & 0.012 & 0.051 & 0.050 & 0.054 \\
 & 200 & 200 & 0.013 & 0.015 & 0.013 & 0.053 & 0.051 & 0.058 \\
\hline
\multirow{5}{*}{\parbox[c]{12em}{\includegraphics[scale=0.27]{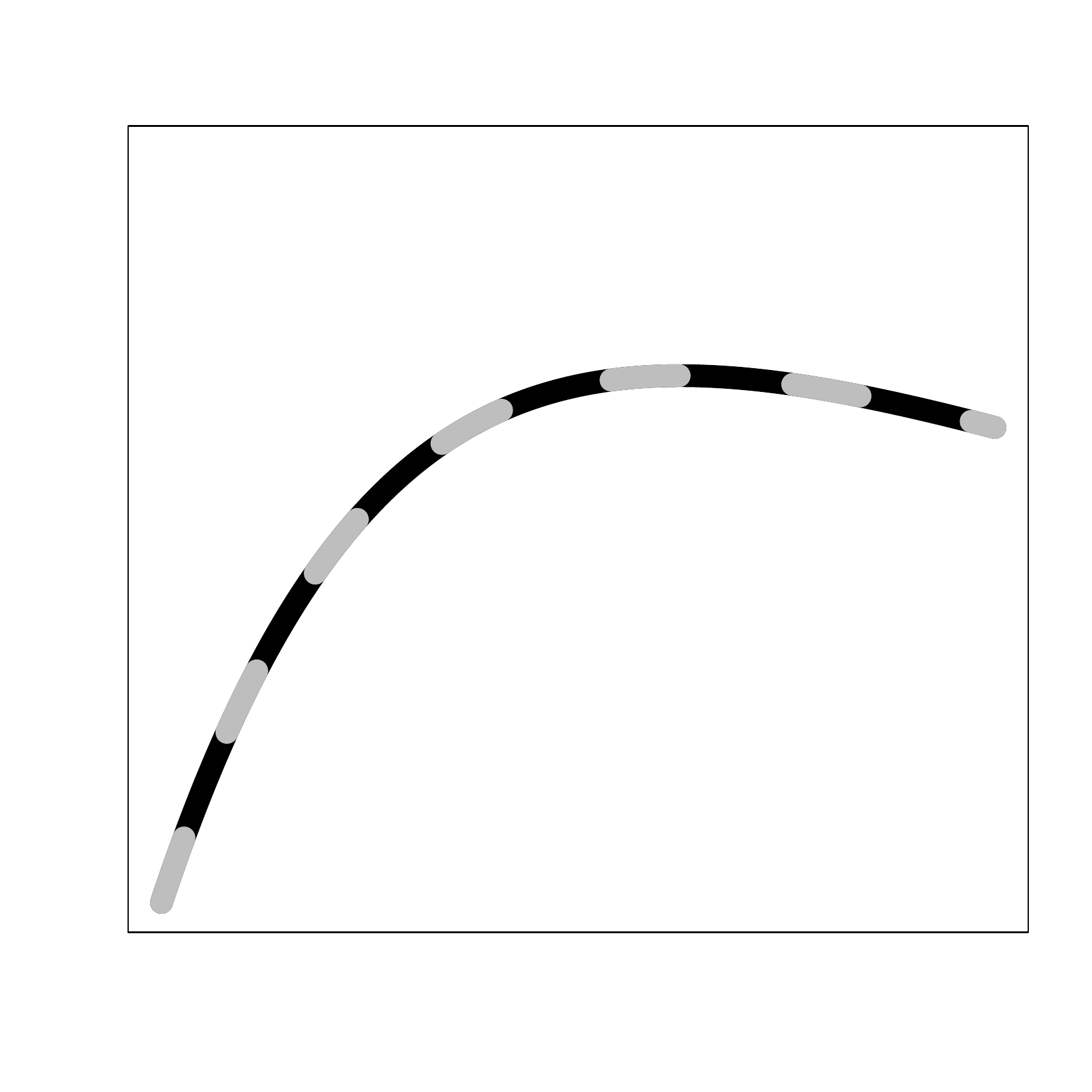}}} & 50 & 50 & 0.012 & 0.016 & 0.016 & 0.064 & 0.062 & 0.076 \\
& 100 & 50 & 0.012 & 0.012 & 0.013 & 0.060 & 0.059 & 0.067 \\
& 100 & 100 & 0.010 & 0.013 & 0.016 & 0.058 & 0.056 & 0.051 \\
& 200 & 100 & 0.007 & 0.008 & 0.010 & 0.044 & 0.054 & 0.058 \\
& 200 & 200 & 0.010 & 0.011 & 0.018 & 0.049 & 0.051 & 0.054 \\
\hline
\end{tabular}
\end{table}

\begin{table}
\caption{Simulation results about empirical power levels for the linear test (Linear), the $L^2$-norm-based test ($L^2$), and the Kolmogorov--Smirnov-type test (KS) under simulation scenaria 5 and 6.}\label{H1a}
\begin{tabular}{c c c c c c c c c}
\hline
&&&\multicolumn{3}{c}{$\alpha=0.01$} & \multicolumn{3}{c}{$\alpha=0.05$} \\
$H_1$ scenario & $n_1$ & $n_2$ & Linear & $L^2$ & KS & Linear & $L^2$ & KS \\
\hline
\multirow{5}{*}{\parbox[c]{12em}{\includegraphics[scale=0.27]{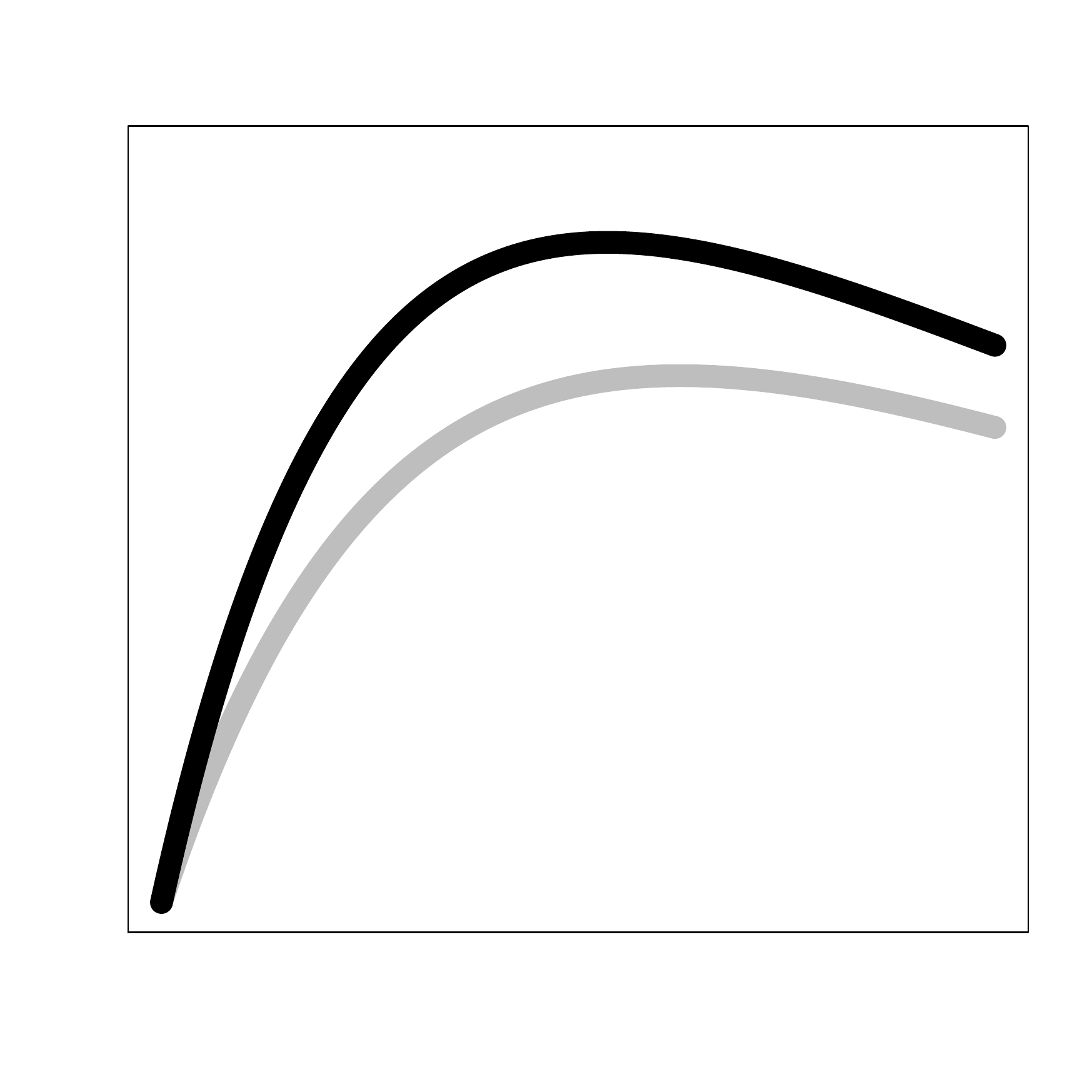}}} & 50 & 50 & 0.074 & 0.069 & 0.055 & 0.170 & 0.162 & 0.166 \\
 & 100 & 50 & 0.085 & 0.084 & 0.075 & 0.210 & 0.200 & 0.205 \\
 & 100 & 100 & 0.106 & 0.101 & 0.101 & 0.251 & 0.240 & 0.240 \\
 & 200 & 100 & 0.143 & 0.142 & 0.127 & 0.307 & 0.296 & 0.307 \\
 & 200 & 200 & 0.233 & 0.206 & 0.207 & 0.448 & 0.417 & 0.397 \\
\hline
\multirow{5}{*}{\parbox[c]{12em}{\includegraphics[scale=0.27]{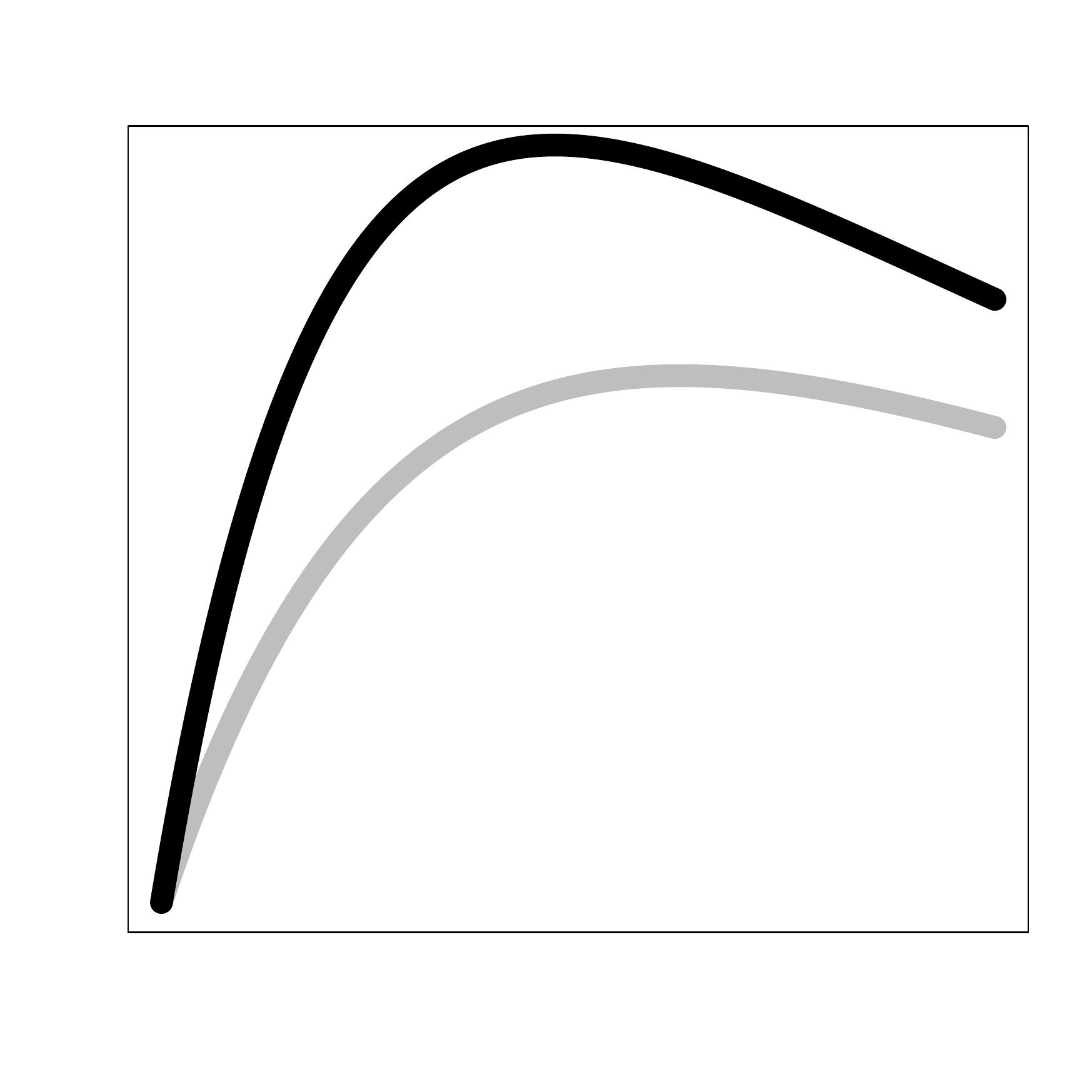}}} & 50 & 50 & 0.183 & 0.177 & 0.178 & 0.361 & 0.348 & 0.370 \\
& 100 & 50 & 0.246 & 0.230 & 0.228 & 0.442 & 0.438 & 0.445 \\
& 100 & 100 & 0.341 & 0.331 & 0.328 & 0.556 & 0.549 & 0.564 \\
& 200 & 100 & 0.458 & 0.464 & 0.480 & 0.703 & 0.696 & 0.708 \\
& 200 & 200 & 0.665 & 0.677 & 0.665 & 0.847 & 0.861 & 0.875 \\
\hline
\end{tabular}
\end{table}

\begin{table}
\caption{Simulation results about empirical rejection rates for the linear test (Linear), the $L^2$ distance test ($L^2$), and the Kolmogorov--Smirnov-type test (KS) under simulation scenaria 7 and 8}\label{H1b}
\begin{tabular}{c c c c c c c c c}
\hline
&&&\multicolumn{3}{c}{$\alpha=0.01$} & \multicolumn{3}{c}{$\alpha=0.05$} \\
$H_1$ scenario & $n_1$ & $n_2$ & Linear & $L^2$ & KS & Linear & $L^2$ & KS \\
\hline
\multirow{5}{*}{\parbox[c]{12em}{\includegraphics[scale=0.27]{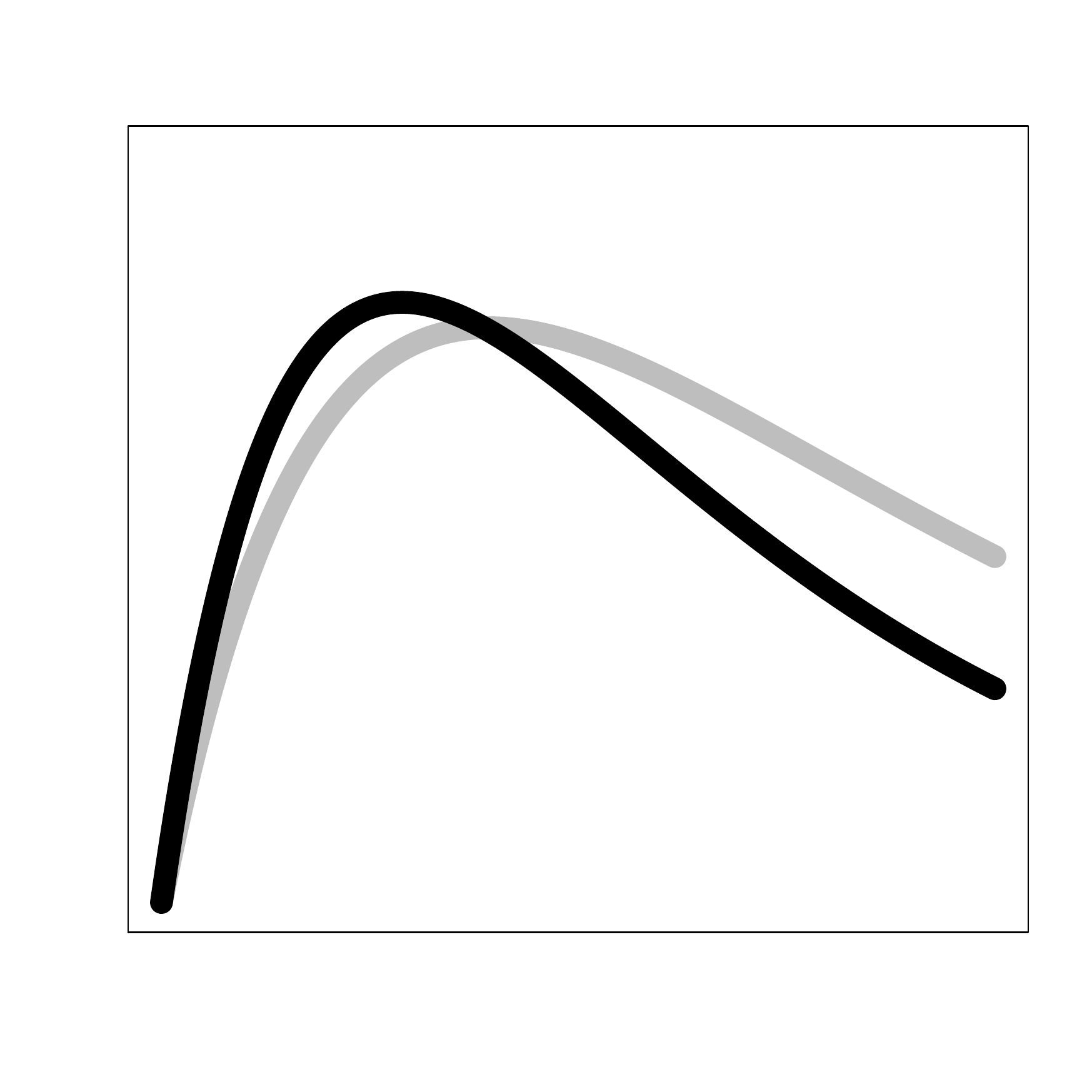}}} & 50 & 50 & 0.015 & 0.033 & 0.033 & 0.064 & 0.099 & 0.123 \\
 & 100 & 50 & 0.016 & 0.028 & 0.033 & 0.064 & 0.111 & 0.135 \\
 & 100 & 100 & 0.017 & 0.056 & 0.060 & 0.074 & 0.164 & 0.169 \\
 & 200 & 100 & 0.018 & 0.066 & 0.058 & 0.094 & 0.204 & 0.203 \\
 & 200 & 200 & 0.026 & 0.102 & 0.105 & 0.095 & 0.304 & 0.288 \\
\hline
\multirow{5}{*}{\parbox[c]{12em}{\includegraphics[scale=0.27]{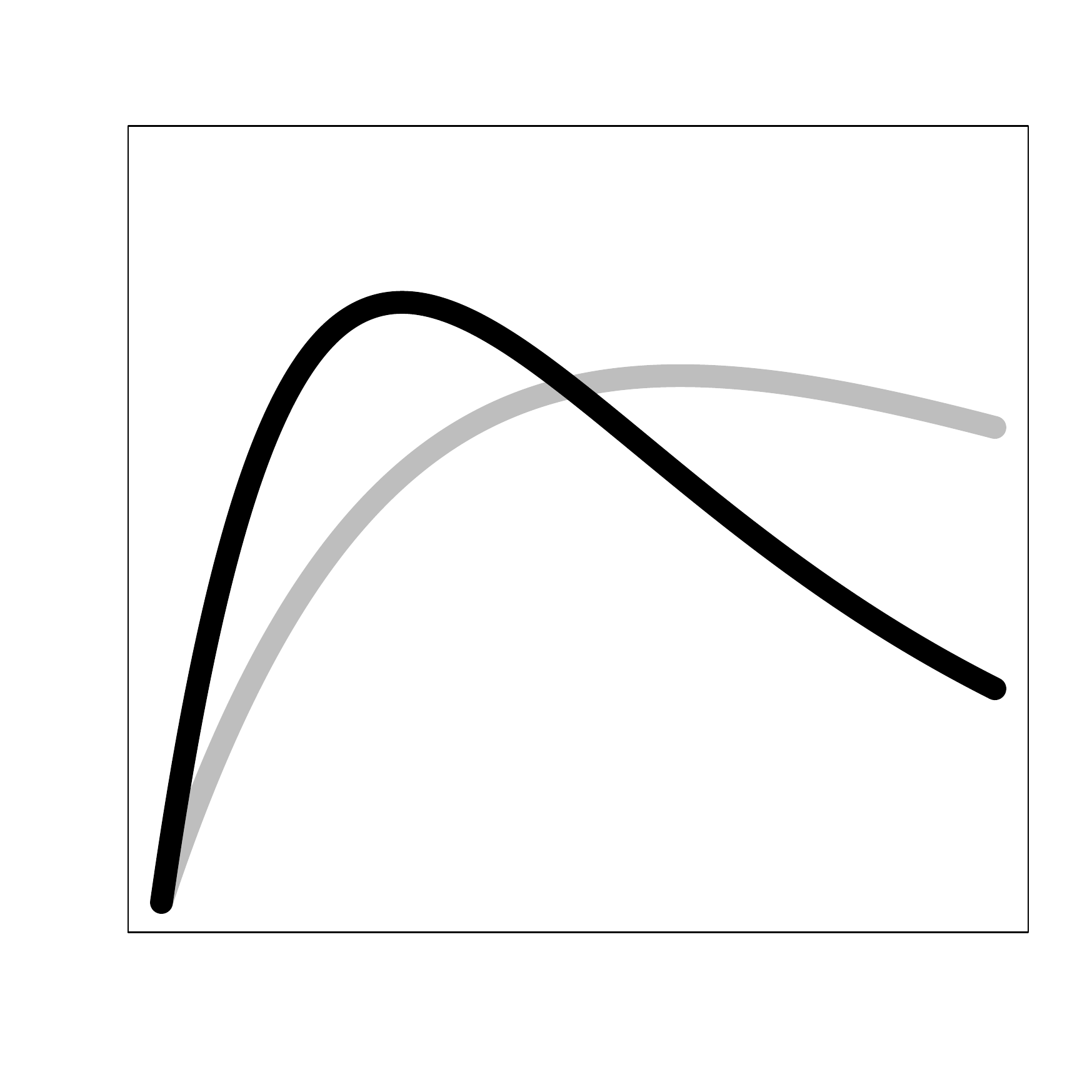}}} & 50 & 50 & 0.016 & 0.055 & 0.138 & 0.055 & 0.247 & 0.356 \\
& 100 & 50 & 0.018 & 0.091 & 0.189 & 0.064 & 0.326 & 0.441 \\
& 100 & 100 & 0.010 & 0.205 & 0.325 & 0.049 & 0.554 & 0.598 \\
& 200 & 100 & 0.016 & 0.351 & 0.490 & 0.062 & 0.712 & 0.795 \\
& 200 & 200 & 0.007 & 0.658 & 0.743 & 0.057 & 0.919 & 0.927 \\
\hline
\end{tabular}
\end{table}

\section{Data analysis}
\label{analysis}
In this section we analyze the data on treatment of early breast cancer from the European Organization for Research and Treatment of Cancer (EORTC) trial 10854. This randomized clinical trial was conducted to evaluate whether the combination of surgery with polychemotherapy is benefical to early breast cancer patients compared to surgery alone. The original analysis of this clinical trial was presented in \citet{EORTC}. 

\begin{figure}
\centering
\includegraphics[scale=0.7]{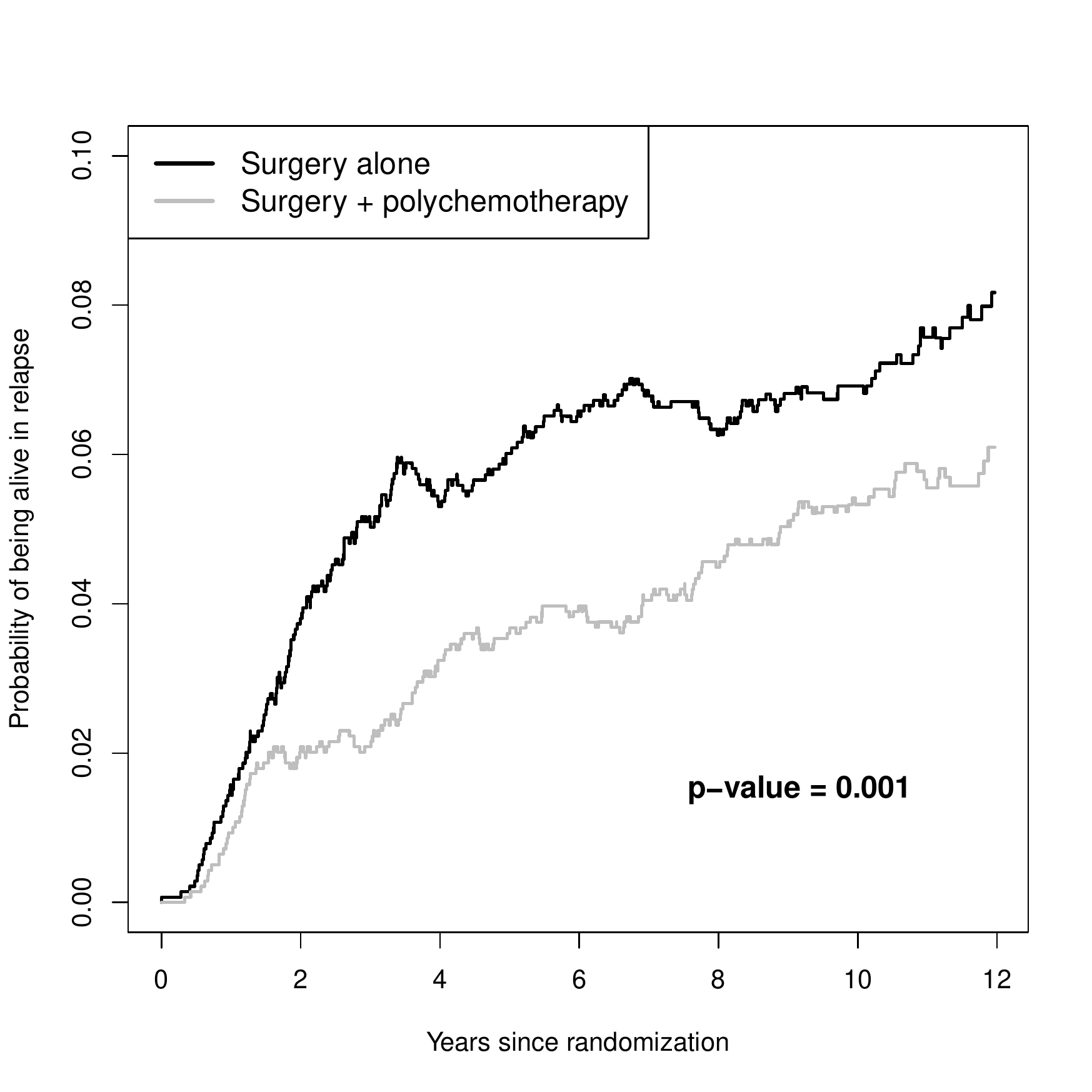}
\caption{\label{dat}Transition probabilities of being alive in relapse by intervention group in the EORTC Trial 10854.}
\end{figure}

In this trial, 1619 patients where randomly assinged to the surgery group and 1559 to the surgery plus polychemotherapy group. The data set contains information about the time to cancer relapse or death. Therefore, an illness-death model is a natural choice for this data set. It is important to note that the transition probability to relapse, which was not analyzed in the original analysis of this trial, is a non-monotonic function of time as patients can move to the ``death'' state after relapse. Thus, standard survival and competing risks analysis methods are not applicable for this transition probability. Here, we focus on this probability which can be interpreted as the probability of being alive and in relapse. The estimated transition probabilities of relapse in the two intervention groups are presented in Figure~\ref{dat}. Based on Figure~\ref{dat}, the probability of being alive and in relapse was lower in the group that received polychemotherapy during surgery. To perform hypothesis testing here we considered the weight function $W_{12}(t)=1$. For the $L^2$-norm-based and Kolmogorov--Smirnov-type tests we considered 1,000 standard normal simulation realizations. The $p$-value from the linear test was 0.001, while the $p$-values from the $L^2$-norm-based and Kolmogorov--Smirnov-type tests were $<$0.001 and 0.004, respectively. These results provide evidence for the superiority of the surgery plus polychemotherapy combination with respect to the transition probability of relapse, in early breast cancer patients.

\section{Concluding remarks}
\label{conclusion}
This paper addressed the issue of direct nonparametric two-sample comparison of transition probabilities $P_{hj}(s,\cdot)$, $s\in[0,\tau)$, for a particular transition $h\rightarrow j$ in a continuous time nonhomogeneous Markov process with a finite state space. The proposed tests were a linear nonparametric test, an $L^2$-norm-based test and a Kolmogorov--Smirnov-type test. Rigorous approaches to evaluate the significance level grounded on modern empirical process theory were provided. Moreover, the $L^2$-norm-based and Kolmogorov--Smirnov-type tests were argued to be consistent against any fixed alternative hypothesis. We also considered extensions of the tests to more complex situations such as cases with missing absorbing states \citep{Bakoyannis19} and non-Markov processes \citep{Putter18}. The simulation study provided numerical evidence for the validity of the proposed testing procedures, which exhibited good performance even with small sample sizes. Finally, a data analysis of a clinical trial on early breast cancer illustrated the utility of the proposed tests in practice.

The issue of nonparametric comparison of transition probabilities in general nonhomogeneous Markov processes has received little attention in the literature. To the best of our knowledge, the only fully nonparametric approach for comparing the transitions probabilitis for a particular transtion in general non-homogeneous Markov processes is a graphical procedure proposed by \citet{Bluhmki18}. This proposal is based on the construction of a simultaneous confidence band for the difference between the transition probabilities of two groups. However, this approach does not provide the exact level of statistical significance. Also, the justification of this approach was based on counting process theory arguments and not on modern empirical process theory. A concequence of that is that this approach cannot be directly adapted to more complex settings that are frequently occur in practice, such as cases with missing absorbing states. An important reason for this is that with more complex estimators, certain predictability conditions assumed by counting process and martingale theory techniques are violated. On the contrary, our proposed methods can be trivially adapted to many other complex settings, provided that appropriate estimators, in the sense of conditions D1--D3, of the transition probabilities exist. Such adaptations can be theoretically justified using the Theorems 3 and 4 provided in our manuscript.

The proposed tests can be easily adapted for the comparison of state occupation probabilities $\Pr(X(t)=j)\equiv P_j(t)=\sum_{h\in\mathcal{I}}P_h(0)P_{hj}(0,t)$, as these are simple linear combinations of the transition probabilities. The state occupation probabilities describe the marginal behavior, i.e. unconditional on the prior history, of the processes and are of interest in many applications, such as in HIV studies focusing on the event history of patients in HIV care \citep{Lee18}. It is important to note that these probabilities can be consistently estimated using the Aalen--Johansen estimator even in non-Markov processes \citep{Datta01}. It is not hard to justify conditions D1--D3 for the state occupation probabiltities under a set of weak regularity conditions. Thus, Theorems 3 and 4 provide a rigorous justification about the use of the proposed tests for comparing state occupation probabilities.

\section*{Acknowledgement}
This project was supported, in part, by the Indiana Clinical and Translational Sciences Institute funded, in part by Grant Number UL1TR002529 from the National Institutes of Health, National Center for Advancing Translational Sciences, Clinical and Translational Sciences Award. We would like to thank the European Organisation for Research and Treatment of Cancer (EORTC) for sharing with us the data from the EORTC trial 10854. The content of this manuscript is solely the responsibility of the authors and does not necessarily represent the official views of the National Institutes of Health and the EORTC.

\bibliographystyle{chicago}
\bibliography{references}

\appendix
\section{Outlines of proofs}
Outlines of the proofs of Theorems 1 and 2 are provided below. The proofs of Theorems 3 and 4 follow from similar arguments and, therefore, are omitted. The proofs rely on empirical process theory techniques \citep{Van96,Kosorok08}. Before providing the proofs it is useful to introduce some notation. First, let $\mathcal{O}$ be the sample space, and $O$ an arbitraty sample point in $\mathcal{O}$. Now, define $\mathbb{P}_nf=\frac{1}{n}\sum_{i=1}^nf(O_i)$, for some measurable function $f:\mathcal{O}\mapsto\mathbb{R}$. Also, define $Pf=\int_{\mathcal{O}}fdP$ to be the expectation of $f$ under the probability measure $P$ on the measurable space $(\mathcal{O},\mathcal{A})$, where $\mathcal{A}$ is a $\sigma$-algebra on $\mathcal{O}$. For simplicity, but without loss of generality, we set the starting point $s=0$ in the following proofs. It has to be noted that conditions C1 and C3--C5 imply the uniform consistency of the standard Aalen--Johansen estimator. This can be shown using similar arguments to those used in the proof of Theorem 1 in \citet{Bakoyannis19}.

\subsection{Proof of theorem 1}
Clearly, Theorem 1 relies on the asymptotic linearity of the estimators $\hat{P}_{n_p,hj}^{(p)}(0,t)$, $p=1,2$. This can be established by first showing the asymptotic linearity of the Nelson--Aalen estimators of the cumulative transition intensities and then by utilizing the functional delta method \citep{Van00}. The steps to achieve this utilize conditions C1 and C3--C5 and arguments similar to those used in the proof of Theorem 2 of \citet{Bakoyannis19}. After this analysis it can be shown that
\[
\sqrt{n_p}[\hat{P}_{n_p,hj}^{(p)}(0,t)-P_{0,hj}^{(p)}(0,t)]=\sqrt{n_p}\mathbb{P}_{n_p}\gamma_{hj}^{(p)}(0,t)+o_p(1), \ \ p=1,2, \ \ h,j\in\mathcal{I},
\]
with 
\[
\gamma_{ihj}^{(p)}(0,t)=\sum_{l\notin\mathcal{T}}\sum_{m\in\mathcal{I}}\int_0^t\frac{P_{0,hl}^{(p)}(0,u-)P_{0,mj}^{(p)}(u,t)}{PY_l^{(p)}(u)}dM_{ilm}^{(p)}(u), \ \ \ \ t\in[0,\tau], \ \ p=1,2.
\]
By Lemma 1 in the supplementary material of \citet{Bakoyannis19} and arguments similar to those use in the proof of Theorem 2 of the same source, it follows that the influence functions $\gamma_{ihj}^{(p)}(0,t)$, $p=1,2$, belong to $P$-Donsker classes of functions. Now, it is not hard to see that under the null hypothesis and by conditions C2 and C6
\begin{eqnarray*}
\sqrt{\frac{n_1n_2}{n_1+n_2}}Z_{hj}&=&\sqrt{1-\lambda}\sqrt{n_1}\mathbb{P}_{n_1}\int_{(0,\tau]}W_{hj}(t)\gamma_{hj}^{(1)}(0,t)dm(t) \\
&&-\sqrt{\lambda}\sqrt{n_2}\mathbb{P}_{n_2}\int_{(0,\tau]}W_{hj}(t)\gamma_{hj}^{(2)}(0,t)dm(t)+o_p(1).
\end{eqnarray*}
Finally, the statement of Theorem 1 follows as a result of the usual central limit theorem and the independence between the two terms, as a consequence of the fact that the two samples are independent.

\subsection{Proof of theorem 2}
Due to the asymptotic linearity of the transition probability estimators $\hat{P}_{n_p,hj}^{(p)}(0,t)$, for $p=1,2$, as argued in the proof of Theorem 1, along with conditions C2 and C6, it follows that
\begin{eqnarray*}
\sqrt{\frac{n_1n_2}{n_1+n_2}}D_{hj}(t)&=&\sqrt{1-\lambda}\sqrt{n_1}\mathbb{P}_{n_1}W_{hj}(t)\gamma_{ihj}^{(1)}(0,t) \\
&&-\sqrt{\lambda}\sqrt{n_2}\mathbb{P}_{n_2}W_{hj}(t)\gamma_{ihj}^{(2)}(0,t)+o_p(1).
\end{eqnarray*}
Now, by the Donsker property of the class of functions $\{\gamma_{hj}^{(p)}(0,t):p=1,2,t\in[0,\tau]\}$ and the uniform boundedness of the class of fixed functions $\{W_{hj}(t):t\in[0,\tau]\}$, it follows that $\{W_{hj}(t)\gamma_{hj}^{(p)}(0,t):p=1,2,t\in[0,\tau]\}$ is also a $P$-Donsker class. Therefore, by the independence between the two samples, it follows that
\[
\sqrt{\frac{n_1n_2}{n_1+n_2}}D_{hj}\leadsto\sqrt{1-\lambda}\mathbb{G}_{1hj}-\sqrt{\lambda}\mathbb{G}_{2hj},
\]
where $\mathbb{G}_{1hj}$ and $\mathbb{G}_{2hj}$ are two independent tight zero-mean Gaussian processes with covariance functions
\[
\sigma_p(v,t)=P[W_{hs}(v)\gamma_{hj}^{(p)}(0,v)][W_{hj}(t)\gamma_{hj}^{(p)}(0,t)], \ \ \ \ p=1,2.
\]
Now, define 
\begin{eqnarray*}
\bar{B}_{hj}(t)&=&\sqrt{1-\lambda}\sqrt{n_1}\mathbb{P}_{n_1}W_{hj}(t)\gamma_{hj}^{(1)}(0,t)\xi^{(1)} \\
&&-\sqrt{\lambda}\sqrt{n_2}\mathbb{P}_{n_2}W_{hj}(t)\gamma_{hj}^{(2)}(0,t)\xi^{(2)},
\end{eqnarray*}
where $\xi^{(p)}$, $p=1,2$, are independent random draws from $N(0,1)$. By the Donsker property of the class $\{W_{hj}(t)\gamma_{hj}^{(p)}(0,t):p=1,2,t\in[0,\tau]\}$, for $h,j\in\mathcal{I}$, and the conditional multiplier central limit theorem \citep{Van96} it follows that 
\[
\sqrt{n_p}\mathbb{P}_{n_p}W_{hj}(\cdot)\gamma_{hj}^{(p)}(0,\cdot)\xi^{(p)}\leadsto \mathbb{G}_{phj}(\cdot),
\]
conditionally on the observed data. Therefore
\[
\bar{B}_{hj}\leadsto\sqrt{1-\lambda}\mathbb{G}_{1hj}-\sqrt{\lambda}\mathbb{G}_{2hj},
\]
conditionally on the observed data. Now it remains to argue that $\sup_{t\in[0,\tau]}|\hat{B}_{hj}(t)-\bar{B}_{hj}(t)|\equiv\|\hat{B}_{hj}(t)-\bar{B}_{hj}(t)\|_{\infty}=o_p(1)$, unconditionally on the observed data. By the triangle inequality it follows that
\begin{eqnarray*}
\|\hat{B}_{hj}(t)-\bar{B}_{hj}(t)\|_{\infty}&\leq&\sqrt{1-\lambda}\left\|\sqrt{n_1}\mathbb{P}_{n_1}[\hat{W}_{hj}(t)\hat{\gamma}_{hj}^{(1)}(0,t)-W_{hj}(t)\gamma_{hj}^{(1)}(0,t)]\xi^{(1)}\right\|_{\infty} \\
&&+\sqrt{\lambda}\left\|\sqrt{n_2}\mathbb{P}_{n_2}[\hat{W}_{hj}(t)\hat{\gamma}_{hj}^{(2)}(0,t)-W_{hj}(t)\gamma_{hj}^{(2)}(0,t)]\xi^{(2)} \right\|_{\infty}.
\end{eqnarray*}
By similar calculations to those in the proof of Theorem 2 in \citet{Bakoyannis19} and conditions C1-C6 it follows that both normed terms in right side the above inequality are $o_p(1)$. This concludes the proof of Theorem 2.

\end{document}